\documentstyle[aps,manuscript]{revtex}

\begin{document}

\title {Mixed population Minority Game with generalized strategies}
\author{P. Jefferies, M. Hart and N.F. Johnson}
\address {Physics Department, Oxford University, Oxford, OX1 3PU, U.K.}
\author{P.M. Hui}
\address {Department of Physics, The Chinese University of Hong Kong, Shatin,
\\ New Territories, Hong Kong}

\maketitle

\begin{abstract} We present a quantitative theory, based on crowd effects, for the
market volatility in a Minority Game played by a mixed population. Below a critical
concentration  of generalized strategy players, we find that the volatility in the
crowded regime remains above the random coin-toss value regardless of the
`temperature' controlling strategy use. Our theory yields good agreement with
numerical simulations.

\end{abstract}
\bigskip


\newpage

Challet and Zhang's Minority Game (MG) offers a simple paradigm in the study
of complex adaptive systems such
as financial markets\cite{challet,challet2,savit,dHR,crowd,us,dublin,sherrington}.
In the MG an odd number
$N$ of agents, each with $s$ strategies and a memory of size $m$, repeatedly
compete to be in the minority. The basic MG features  agents who use their highest
scoring strategy. As pointed out by Marsili {\em et al}
\cite{marsili},  a probabilistic strategy choice reflects a particular behavioral
model and has a long tradition in economics. Cavagna {\em et al} performed
numerical simulations of the MG in which agents use such an exponential probability
weighting controlled by a `temperature' $T$ \cite{sherrington}; this is called the
Thermal Minority Game (TMG) although it has been noted that  
$T^{-1}$ more closely represents the agents' learning rate (see Ref.
\cite{marsili2}). Challet {\em et al}, in addition to presenting a detailed
spin-glass theory for the basic MG\cite{challet2},  have recently identified
problems\cite{comment} with the TMG results of Cavagna et al.
\cite{sherrington}. Our own interest in the TMG has focused on the finding
that the volatility (i.e. standard deviation)
$\sigma$ can be reduced from being larger than
the random coin-toss value (`worse-than-random') to being 
smaller than
the random coin-toss value (`better-than-random')  just by altering the relative
probability weighting\cite{sherrington}. We recently provided an analytic theory
which explains  this effect in terms of crowds\cite{note}. 

In this paper, we consider a generalized Minority Game in which a concentration
$q$ of agents employ such probabilistic strategy selection at each turn of the
game. We present a quantitative theory, based on crowd effects, which yields good
agreement with numerical simulations. We find that below a critical concentration
$q_c^*$, the volatility $\sigma$ remains larger than the
random coin-toss value regardless of the `temperature' $T$ controlling the strategy
selection. 

Our generalized Minority Game contains $N$ agents who choose repeatedly between
option 0 (e.g. buy) and option 1 (e.g. sell). The winners are those in the minority
group, e.g. sellers win if there is an excess of buyers. The outcome at each
timestep represents the winning decision, 0 or 1. A common bit-string of the
$m$ most recent outcomes\cite{memory} is made available to the agents at each
timestep. The agents randomly pick $s$ strategies at the beginning of the game,
with repetitions allowed, from the pool of all possible strategies. We focus on
$s=2$.   After each turn, the agent assigns one (virtual) point to each of his
strategies which would have predicted the correct outcome. In the basic MG, each
agent plays the most successful strategy in his possession, i.e. the one with the
most virtual points. Here we instead allow a concentration $q$ of agents to follow
a more general behavioral model: in particular, these agents play their worst
strategy with probability
$\theta$, and hence play their best strategy with probability
$(1-\theta)$. These $qN$ agents will be called `TMG agents' because of the direct
connection with the Thermal Minority Game\cite{sherrington,tmg}. The remaining
$(1-q)N$ agents choose their best strategy with probability unity (i.e.
$\theta=0$ as in the basic MG)  hence they will be called `MG agents'.  

Figure 1  shows the volatility $\sigma$ obtained from numerical
simulations of a game with $N=101$ and $m=2$, as a function of $q$ at various fixed
$\theta$ values.  The dashed line shows the random coin-toss
value for $N$ agents, given by ${\sqrt N}/2$. 
Figure
2 shows an example of the corresponding numerical results for $\sigma$
as a function of $\theta$ at fixed $q$. 
A definite trend can be seen in Figs. 1 and 2, despite the numerical spread which
arises naturally for different runs: As the
concentration 
$q$ of TMG agents increases, or the probability $\theta$ (i.e. $T$
\cite{tmg}) increases, the 
volatility $\sigma$ decreases. At $q=1$ (Fig. 2) we reproduce the main finding of
Ref.
\cite{sherrington} whereby $\sigma$ falls from worse-than-random to
better-than-random with increasing $\theta$ (`temperature' $T$\cite{tmg}). The
numerical results in Fig. 1 indicate that below a critical $q$, 
$\sigma$ lies in the worse-than-random regime regardless of
$T$. Our goal is to develop a quantitative theory describing the trend in the
run-averaged  volatility (i.e. the volatility averaged over
initial strategy configurations) as a function of
$q$ and
$\theta$.

In Ref.
\cite{us} we  presented a quantitative theory for  the volatility $\sigma$ in the
basic MG which yields good agreement with numerical simulations over the entire
parameter range of interest. The theory is based on the consideration of the
combined actions of crowds and their anticorrelated partners (anticrowds). For each
crowd-anticrowd pair, the action of the anticrowd will effectively nullify the
action of the crowd if they are of similar size,  hence reducing the volatility
$\sigma$ \cite{crowd,us}.   For small $m$ and large $N$ \cite{us} the crowds are
typically much larger than the anticrowds\cite{us} hence the basic MG is in the
`crowded' regime (i.e. 
$\sigma$ is larger than the random coin-toss value); this is the regime of interest
here since we are focusing on the transition of $\sigma$ from worse-than-random to
better-than-random. A cruder version of our crowd theory was earlier shown to
provide a good quantitative description for the MG played by a population of
mixed-memory agents\cite{dublin,memory}. Given this success, we build the
present theory using the same crowd-anticrowd ideas. Consider any two strategies
$r$ and
$r^*$ within the list of $2^{m+1}$ strategies in the reduced strategy space
\cite{challet,us}. At any moment in the game, the strategies can be ranked
according to their virtual points,
$r=1,2
\dots 2^{m+1}$ where $r=1$ is the best strategy, $r=2$ is second best, etc. Note
that in the small $m$ regime of interest, the strategy ranking in order of
decreasing virtual points can be taken to be identical to the strategy ranking in
order of decreasing number of users (i.e. decreasing popularity) to a good
approximation
\cite{us}. Accidental degeneracies may arise whereby two different
strategies momentarily have identical virtual points, however these degeneracies are
removed when considering an average over several timesteps - hence any agent
holding two strategies with the same ranking must necessarily have picked the same
strategy twice. Let
$p(r,r^*|r^*\geq r)$ be the probability that a given agent picks $r$ and
$r^*$, where $r^*\geq r$.  Let
$p(r,r^*|r^*\leq r)$ be the probability that a given agent picks $r$ and
$r^*$, where $r^*\leq r$. The probability that a TMG agent plays
$r$ is given by
\begin{eqnarray} p_r^{\rm TMG} & = & \sum_{r^*=1}^{2^{m+1}} [\ \theta\ 
p(r,r^*|r^*\leq r) + \ (1-\theta)
\ p(r,r^*|r^*\geq r)]\nonumber \\ & = & 
 \ \theta\ p_-(r)  + 2^{-2(m+1)}\ \theta + \ (1-\theta)\  p_+(r)
\end{eqnarray} where $p_+(r)=\sum_{r^*} p(r,r^*|r^*\geq r)$ is the probability that
the agent has picked
$r$ {\em and} that $r$ is the agent's best (or equal best) strategy; 
$p_-(r)=\sum_{r^*} p(r,r^*|r^*<r)$ is the probability that the agent has picked
$r$ {\em and} that $r$ is the agent's worst strategy. The factor $2^{-2(m+1)}$
in Eq. (1) originates from $p(r,r^*|r^*=r)$. The
probability that an MG agent plays
$r$ is given by
\begin{equation} p_r^{\rm MG} = p_+(r) \ \ .
\end{equation} It follows that 
$p_+(r) + p_-(r) = p(r)$ where 
\begin{equation} p(r) = 2^{-m} ( 1 - 2^{-(m+2)})
\end{equation} is the probability that an agent holds strategy
$r$ after his $s=2$ picks with no condition on whether it is best  or worst.    Now
we consider the mean number of agents $n_r$ playing strategy $r$ in the
mixed-population game containing a concentration $q$ of TMG agents and $(1-q)$ of
MG agents. This is given  by
\begin{eqnarray} n_r & = &  \ q\  N\  p_r^{\rm TMG} + \ (1-q)\  N\  p_r^{\rm MG}
\nonumber \\ & = & \ N\ (1-2\ q\ \theta)\ p_+(r) + N\ q\ \theta\  p(r)
+ 2^{-2(m+1)}\ N\ q\ \theta \
\ .
\end{eqnarray}  If ${n}_r$ agents all use strategy $r$, they will act
as a `crowd', i.e. they make the same decision. If
${n}_{\bar r}$ agents simultaneously use the strategy
$\bar r$ anticorrelated to $r$,  they will make the  opposite (anticorrelated)
decision and hence act as an `anticrowd' \cite{us}. The standard deviation
$\sigma(q,\theta)$ in the  number of agents making a particular decision (say 0)
is given by\cite{us}
\begin{equation}
\sigma(q,\theta) =
\bigg[ \frac{1}{2} \sum_{r=1}^{2^{m+1}}
\frac{1}{4}|n_r-n_{\bar r}|^2 \bigg]^{\frac{1}{2}}\ \ .
\end{equation} Using Eqs. (3), (4) and (5)  for $r$ and ${\bar r} =2^{m+1} + 1 - r$
we obtain
\begin{equation}
\sigma(q,\theta) = [1-2\ q\ \theta] \ \{\sigma(q,\theta)\}_{q\theta=0}
\end{equation} where
$\{\sigma(q,\theta)\}_{q\theta=0}$ is just the standard deviation for the basic MG
(i.e. $q=0$ and/or $\theta=0$). In Ref. \cite{us}, we provided an analytic
formulation of  
$\{\sigma(q,\theta)\}_{q\theta=0}$.  However, Eq. (6) is more general in that it
does not specify the level of approximation used to obtain
$\{\sigma(q,\theta)\}_{q\theta=0}$. 

Our theory (Eq. (6)) predicts that the effect
on the volatility caused by a change in population
composition  and/or `temperature' can be described by a simple prefactor $[1-2 q
\theta]$. Provided that the basic MG is in the crowded
regime as discussed earlier, Eq. (6) should hold for all
$N$ and $m$ and hence any value of
$\{\sigma(q,\theta)\}_{q\theta=0}$. Hence we can predict the critical value $q_c$
for fixed $\theta$, or
$\theta_c$ for fixed $q$,  at which
$\sigma(q,\theta)$ crosses from worse-than-random to better-than-random. For a
given value of
$\theta$, it follows from Eq. (6) that 
\begin{equation} q_{c}(\theta) = \frac{1}{2\theta}-
\frac{\sqrt{N}}{4\theta}\frac{1}{\{\sigma(q,\theta)\}_{q\theta=0}}
\ .
\end{equation} A similar expression follows for $\theta_c(q)$.  Given that $0\leq
\theta\leq 1/2$ \cite{tmg}, Eq. (7) implies that the 
run-averaged numerical volatility should lie above the random coin-toss
value if
$q<q_c^*$ where
\begin{equation} q_{c}^* = 1 -
\frac{\sqrt{N}}{2}\frac{1}{\{\sigma(q,\theta)\}_{q\theta=0}}
\ ,
\end{equation} regardless of `temperature' $T$ \cite{tmg}. Since we
are considering $N$ and $m$ values such that the basic MG is in the
worse-than-random regime, $\{\sigma(q,\theta)\}_{q\theta=0}\geq\sqrt{N}/2$ and
therefore $0\leq q_c^*\leq 1$ as required. Similarly $\sigma(q,\theta)$  
will remain
above the random coin-toss value for all $q$ if
$\theta <\theta_c^*$ where
\begin{equation}
\theta_{c}^* = \frac{1}{2} -
\frac{\sqrt{N}}{4}\frac{1}{\{\sigma(q,\theta)\}_{q\theta=0}}
\ .
\end{equation}

Figures 1 and 2 compare the present theory (Eq. (6)) to the
numerical simulations. The theoretical points lie within the numerical spread over
a wide range of
$q$ and
$\theta$ values, and hence provide a quantitative explanation of the observed
trends. Since we are interested in
testing the simple prefactor scaling predicted by Eq. (6),  we have generated Figs.
1 and 2 using the numerical value of
$\{\sigma(q,\theta)\}_{q\theta=0}$ obtained from the basic MG; we emphasize,
however, that an analytic formulation for
$\{\sigma(q,\theta)\}_{q\theta=0}$ is provided in Ref.
\cite {us}.
Although not relevant for the main results of this paper, the present
theory (Eq. (6)) begins to underestimate the numerical results in the
better-than-random regime as
$\sigma(q,\theta) \rightarrow 0$ (not shown). 
There are shortcomings in the theory which can explain this effect; in
particular, $p_r^{\rm TMG}$ in Eq. (1) is an average value over the configuration 
space of possible initial strategy picks, and over time. It has a
decreasing dependence on $r$ as $\theta\rightarrow 0.5$, hence giving rise to
$\sigma=0$ (i.e. exact crowd-anticrowd cancellation) for $q=1$ and $\theta=0.5$.
Consider $q=1$ and $\theta=0.5$; for a particular configuration of
strategies picked at the start of the game, and at a particular moment in time, the
number of agents using each strategy is typically distributed {\em around} the
value $N \ 2^{-(m+1)}$. It is this non-flat distribution describing the strategy-use
by coin-flipping TMG agents which will actually give rise to a non-zero
$\sigma$. 
Having obtained
$\sigma$ for a given initial
configuration of strategies, the average should then be taken
over all initial strategy configurations. We have shown that carrying out this
procedure yields a non-zero theoretical $\sigma$ and restores agreement with the
numerical data in the better-than-random regime\cite{us2}.

 Figure 3 shows the theoretical `phase diagram' for the volatility
$\sigma(q,\theta)$. The curve $q_c(\theta)$, or equivalently
$\theta_c(q)$, separates the regions where $\sigma$
is worse-than-random and better-than-random. Also indicated are $q_c^*$ and
$\theta_c^*$.

In summary we have analyzed a mixed population Minority Game
with generalized strategies. The main feature of the numerical results
regarding volatility-reduction from worse-than-random to better-than-random, can be
explained quantitatively without having to solve the detailed game dynamics. More
generally, it is clear that there will be some properties of MG games which cannot
be described using such time- and configuration-averaged theories as used here (see
Ref.
\cite{marsili2}). Moreover, the volatility in real financial
markets is more likely to correspond to a {\em single} run which evolves from a
specific initial configuration of agents' strategies. Our crowd-anticrowd viewpoint
can, however, be extended to deal with these game dynamics via the
dynamical equations governing the co-evolution of the
crowd-anticrowd populations.  The correct equations are not continuous in
time in general. The MG dynamics described in terms of the time-evolution of 
crowds-anticrowds will be presented elsewhere\cite{us2}.

\newpage 
\centerline{\bf Figure Captions}

\bigskip

\noindent Figure 1: Comparison between numerical simulations (circles) and the
present theory (solid line using Eq. (6)) for the volatility $\sigma$ as a
function of TMG agent concentration
$q$ at fixed $\theta$: (a) $\theta=0.1$, (b)
$\theta=0.3$ and (c) $\theta=0.5$. 
The `temperature' $T$ corresponding to each $\theta$ is given (see Ref.
\cite{tmg}). 
$N=101$ and $m=2$. Numerical data are shown for  several runs. Dashed line shows 
random coin-toss value. 
\bigskip

\noindent Figure 2: Comparison between numerical simulations (circles) and the
present theory (solid line using Eq. (6)) for
the volatility $\sigma$ as a function of the probability $\theta$ for a pure
population of TMG agents (i.e. $q=1$). $N=101$ and
$m=2$. Numerical data are shown for  several runs. Dashed line shows 
random coin-toss value. 
\bigskip

\noindent Figure 3: `Phase diagram' in $(q,\theta)$ space. Curve
corresponds to Eq. (7) and separates regions where  volatility $\sigma$ lies above
the random coin-toss value (`worse-than-random') and below (`better-than-random').
$N$ and $m$ values as in Fig. 1.


\newpage




\begin{thebibliography}{99}
\bibitem{challet} D. Challet and Y.C. Zhang, Physica A {\bf 246}, 407 (1997); {\em
ibid.} {\bf 256}, 514 (1998); {\em ibid.} {\bf 269}, 30 (1999).
\bibitem{challet2} D. Challet and M. Marsili, Phys. Rev. E {\bf 60}, R6271 (1999);
D. Challet, M. Marsili, and R. Zecchina, Phys. Rev. Lett. {\bf 84}, 1824 (2000); D.
Challet and M Marsili, cond-mat/9908480. 
\bibitem{savit} R. Savit, R. Manuca and R. Riolo, Phys. Rev. Lett.  {\bf 82}, 2203
(1999).
\bibitem{dHR} R. D'Hulst and G.J. Rodgers, Physica A {\bf 270}, 514 (1999).
\bibitem{crowd} N.F. Johnson, M. Hart and P.M. Hui, Physica A {\bf 269}, 1 (1999). 
\bibitem{us} M. Hart, P. Jefferies, N.F. Johnson and P.M. Hui, cond-mat/0003486.
\bibitem{dublin} N.F. Johnson, M. Hart, P.M. Hui and D. Zheng, cond-mat/9910072;
N.F. Johnson, P.M. Hui, D. Zheng and M. Hart, J. Phys. A: Math. Gen. {\bf 32} L427
(1999).
\bibitem{sherrington} A. Cavagna, J.P. Garrahan, I. Giardina and D. Sherrington,
Phys. Rev. Lett. {\bf 83}, 4429 (1999); see also J.P. Garrahan, E. Moro and D.
Sherrington, cond-mat/0004277. 
\bibitem{marsili} M. Marsili, D. Challet and R. Zecchina, cond-mat/9908480. 
$T$-dependent, Boltzmann-like strategy weightings were discussed by M. Marsili at
the International Workshop on Econophysics and Statistical Finance (Palermo,
September 1998); see Physica A {\bf 269}, 9 (1999). 
\bibitem{marsili2} M. Marsili and D. Challet, Adv. Complex Systems {\bf 1}, 1
(2000).
\bibitem{comment} D. Challet, M. Marsili and R. Zecchina, cond-mat/0004308. 
 \bibitem{note} M. Hart, P. Jefferies, N.F. Johnson and P.M. Hui, cond-mat/0004063.
\bibitem{memory} See  D. Challet and M. Marsili, cond-mat/0004196 and references
therein for demonstrations confirming the relevance of the actual
memory in the MG, in contrast to the claim of 
A. Cavagna [Phys. Rev. E {\bf 59}, R3783
(1999)].
\bibitem{tmg} The Thermal Minority Game discussed in Ref. \cite{sherrington} depends
on a parameter
$T$ (or equivalently $1/\beta$) called a `temperature'. We could similarly define 
$T$ by setting the probability of playing the worst strategy
$\theta = e^{-\beta}/(e^{\beta}+e^{-\beta})$. Hence $T=2[{\rm ln} (\theta^{-1}
-1)]^{-1}$. $T=0$ corresponds to $\theta=0$ while
$T\rightarrow \infty$ corresponds to $\theta\rightarrow 1/2$, hence we will only
consider $0\leq\theta\leq 1/2$ in this paper.  
\bibitem{us2} M. Hart, P. Jefferies, N.F. Johnson and P.M. Hui, in preparation.


\end{thebibliography}
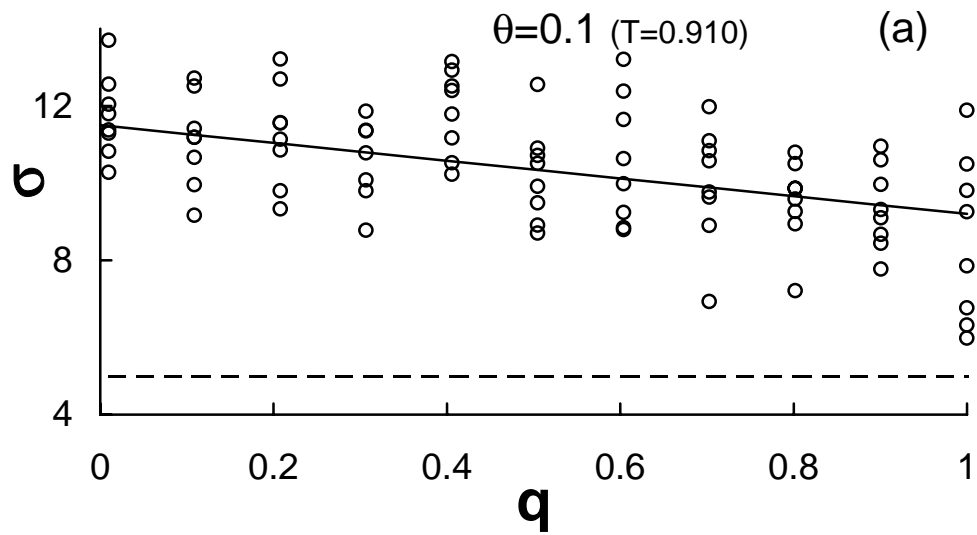
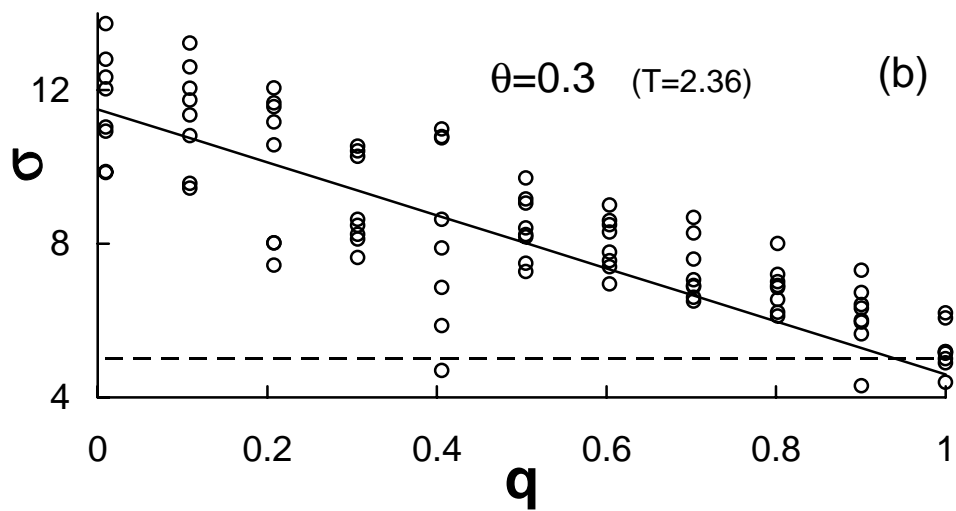
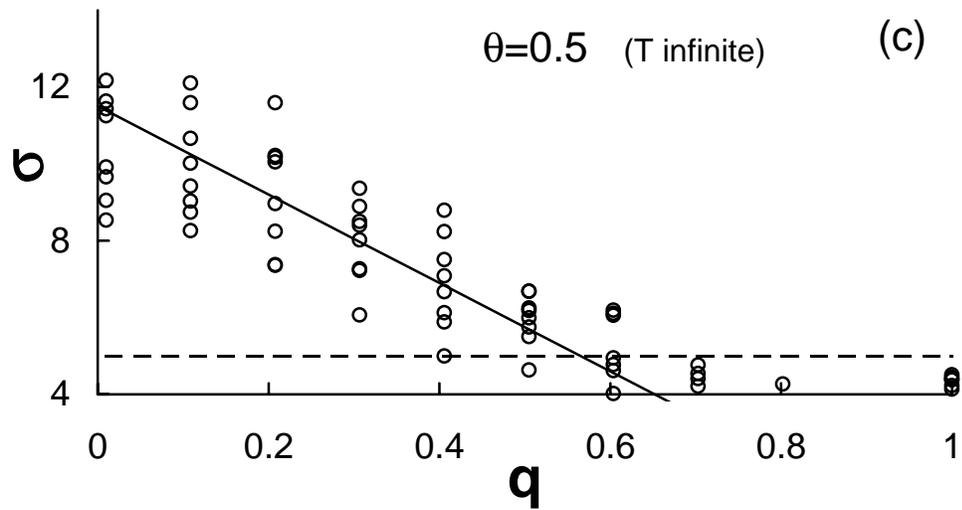
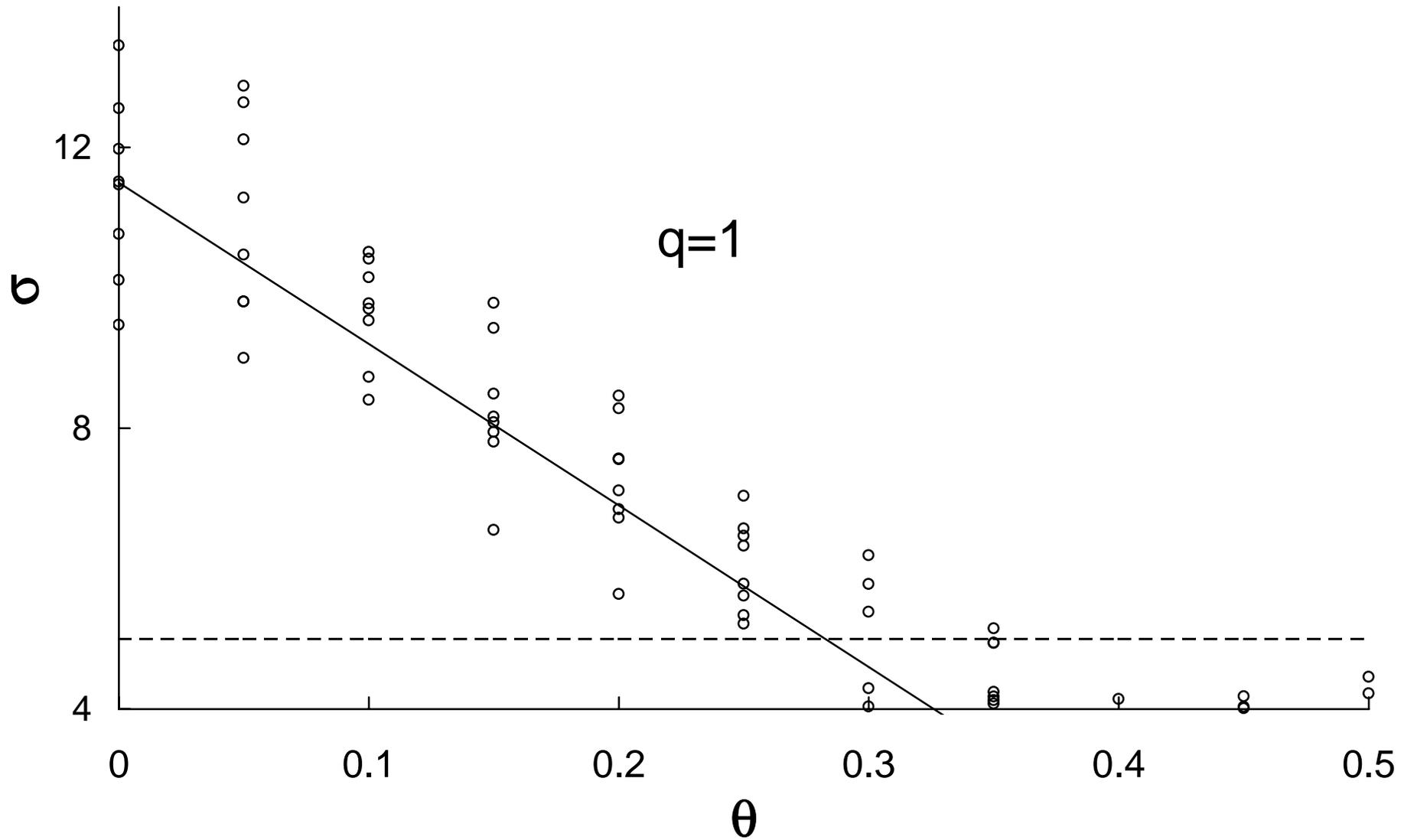
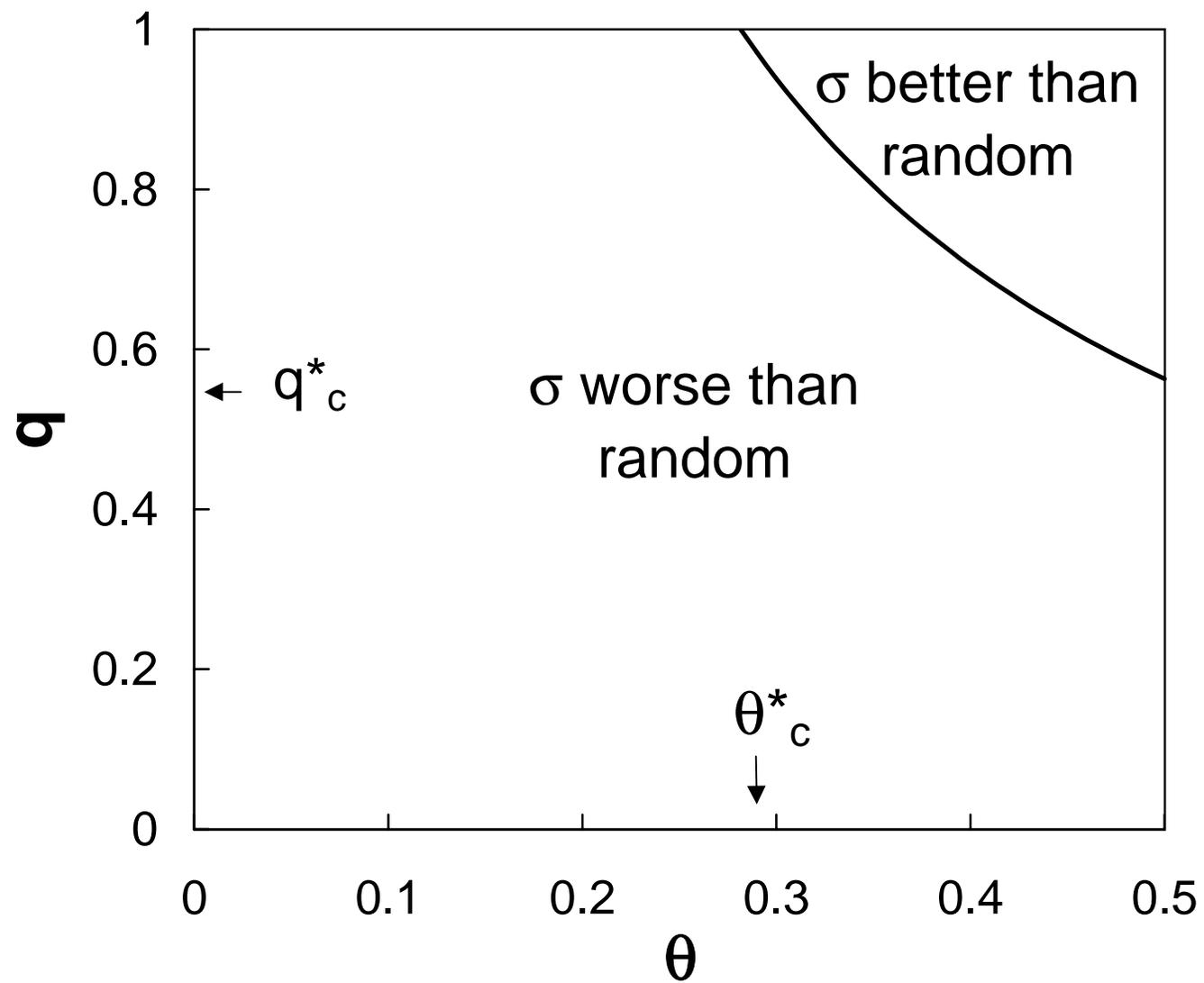
\end{document}